\documentclass[useAMS,usenatbib]{mn2e}

\usepackage{txfonts,graphicx,amssymb,rotating}

\title[RS Oph and Type Ia SNe]{Disc instability in RS Ophiuchi: a path to Type Ia supernovae?}

\author[R.D.Alexander et al.]  {R.D.Alexander$^{1,}$\thanks{email: richard.alexander@leicester.ac.uk}, G.A.Wynn$^1$,
  A.R.King$^1$ and J.E.Pringle$^{1,2}$ \\$^1$ Department of Physics \&
  Astronomy, University of Leicester, Leicester, LE1 7RH\\$^2$
  Institute of Astronomy, Madingley Road, Cambridge, CB3 0HA}
  
\begin{document}
\voffset=-0.25in
\newcommand{\Msunyr}{M$_{\odot}$yr$^{-1}$}
\newcommand{\Msun}{M$_{\odot}$}

\pagerange{\pageref{firstpage}--\pageref{lastpage}} \pubyear{2011}

\date{Accepted 2011 August 17.  Received 2011 August 16; in original form 2011 June 22}

\maketitle

\label{firstpage}

\begin{abstract}
We study the stability of disc accretion in the recurrent nova RS Ophiuchi.  We construct a one-dimensional time-dependent model of the binary-disc system, which includes viscous heating and radiative cooling and a self-consistent treatment of the binary potential.  We find that the extended accretion disc in this system is always unstable to the thermal--viscous instability, and undergoes repeated disc outbursts on $\sim10$--20yr time-scales.  This is similar to the recurrence time-scale of observed outbursts in the RS Oph system, but we show that the disc's accretion luminosity during outburst is insufficient to explain the observed outbursts.  We explore a range of models, and find that in most cases the accretion rate during outbursts reaches or exceeds the critical accretion rate for stable nuclear burning on the white dwarf surface.  Consequently we suggest that a surface nuclear burning triggered by disc instability may be responsible for the observed outbursts.  This allows the white dwarf mass to grow over time, and we suggest that disc instability in RS Oph and similar systems may represent a path to Type Ia supernovae.
\end{abstract}

\begin{keywords}
accretion, accretion discs -- instabilities -- novae, cataclysmic
variables -- binaries: symbiotic -- supernovae: general -- white dwarfs
\end{keywords}


\section{Introduction}

The recurrent novae (RNe) are a small set of close binary systems which are observed to have nova--like outbursts every $\sim 10$--100 years. The main reason for interest in them is their possible relevance in the search for progenitors of Type Ia supernovae (SNe Ia).  It is generally agreed that SNe Ia occur when a white dwarf (WD) is driven over the Chandrasekhar mass $M_{\mathrm {Ch}}$ \citep[e.g.,][]{hn00}. The main obstacle to arranging this is that accretion of hydrogen--rich matter on to a WD generally appears to result in a thermonuclear runaway -- a nova explosion -- once a certain mass has been accreted. All novae recur, and RNe earn their name simply because their recurrence times are short enough to be observed.  However, nova explosions remove most or all of the mass that was accumulated since the last explosion, and it seems likely that the masses of accreting WDs are slowly eroded by this process, rather than increasing \citep[e.g.,][]{nomoto82,yaron05}.  The only exception to this is that the accreting mass can apparently undergo steady nuclear burning if it arrives at a rather narrowly confined rate $\dot M_{\mathrm {steady}} \simeq 3$--$8 \times 10^{-7}$\Msunyr~\citep[e.g.,][]{sb07,nomoto07}.

Two solutions to this problem are currently suggested. One is that SNe Ia result from the merger of two WDs with total mass exceeding $M_{\mathrm {Ch}}$ \citep{webbink84,it84}; the other model picks out an unusual subset of accreting WD systems \citep{wi73,nomoto82}.  Observations of the progenitors of Type Ia SNe are scarce, but generally point towards growth by accretion rather than a double-degenerate origin \citep[e.g.,][]{vn08}.  However, it is far from obvious how this accretion proceeds.  The vast majority of accreting WDs are cataclysmic variables (CVs), where the companion star has a lower mass than the WD. Mass transfer here requires orbital angular momentum loss, or in rare cases nuclear expansion of the companion. However if the close companion is more massive than the WD, mass transfer is self-sustaining on the thermal time-scale $t_{\mathrm {KH}}$ of the companion, a so-called supersoft source. For a likely companion mass $\sim 1$\Msun this mass transfer rate is plausibly close to $\dot M_{\mathrm {steady}}$, allowing the WD mass to grow \citep{lvdh97}. The model requires some fine tuning to ensure that the mass transfer rate remains close to $\dot M_{\mathrm {steady}}$, and can only increase the WD mass at most to the current companion mass. Recently \citet{gb10} and \citet{bg11} have shown that the total soft X-ray emission of both early- and late-type galaxies implies numbers of supersoft sources which are too low to account for the global SNe Ia rate.

The RNe offer intriguing insight into these problems.  Fitting their very short recurrence times into the general picture of classical novae requires the RNe to have WD masses $M_1$ very close to $M_{\rm Ch}$. The argument runs as follows \citep[e.g.,][]{tl86}.  A thermonuclear runaway requires a critical pressure $P_{\rm crit} \sim 2\times 10^{19}~{\rm dyne\, cm^{-2}}$, independent of WD mass, at the base of the accreted layer of mass $\Delta M$ so that
\begin{equation}\label{eq:pcrit}
P_{\rm crit} = {GM_1\Delta M\over R_{\mathrm {wd}}^4}
\end{equation}
where $R_{\mathrm {wd}}$ is the WD radius. The short recurrence time-scale of RNe allows little time to accumulate matter, so $\Delta M$ is unusually small. From Equation \ref{eq:pcrit} we see that $\Delta M \propto R_{\mathrm {wd}}^4$, so recurrent novae require small $R_{\mathrm {wd}}$ and thus WD masses very close to $M_{\mathrm {Ch}}$. More recent work suggests that other factors, such as the composition and thermal profile of the igniting layer, may also affect the critical mass $\Delta M$ \citep[e.g.,][]{sb09,kasliwal11}, but short recurrence time-scales still generally require WD masses close to $M_{\mathrm {Ch}}$.  Unfortunately the difficulty of measuring WD masses in CVs means that there is little strong evidence either for or against this prediction. \citet{thoroughgood01} find $M_1= 1.55 \pm
0.24$\Msun~for U Sco, consistent with the theoretical prediction, but few other dynamical masses have been measured.

However, other interpretations of the recurrent novae have been suggested. In particular, the observation of a radio jet in the recurrent nova RS Oph \citep*{srm08} is difficult to reconcile with the idea of a thermonuclear runaway \citep{kp09}.  Well-collimated jets require the presence of an accretion disc, whose inner parts would be destroyed by a near-spherical nova explosion since the nuclear energy release per gram considerably exceeds the gravitational binding energy at the WD surface \citep*[e.g.,][]{frank02}.  Therefore, at least at the onset of the outburst, thermonuclear energy release cannot dominate the gravitational contribution.

\citet{kp09} noted that the very long orbital period of RS Oph ($\sim 454$~days) and the relatively low mass M-giant secondary imply a wide orbital separation ($\gtrsim 10^{13}$cm) and a very large accretion disc \citep[see also][]{wynn08}.  The outer regions of such a disc are cool, and the disc is therefore subject to the usual thermal--viscous disc instability which powers dwarf novae (see e.g. Lasota, 2001 for a review). \citet{kp09} therefore attempted to interpret the outbursts of RS Oph as very large dwarf nova events, where the outburst luminosity comes solely from gravitational accretion. The aim of this study is to test this
interpretation in detail.

In this paper we present a one-dimensional, time-dependent model for disc accretion in the RS Oph system. Our model includes viscous heating and radiative cooling, as well as a self-consistent treatment of the binary potential, and is able to reproduce outbursts due to the thermal--viscous instability.  We also model the observed emission from the disc, during both quiescence and outburst.  The disc is fed by Roche lobe overflow from the giant secondary, and we find that the disc is always unstable to the thermal--viscous instability.
The outbursts naturally recur on $\sim 10$yr time-scales, but the accretion disc is not sufficiently luminous to explain the observed outbursts.  The total mass accreted during these outbursts can be large, $\lesssim 10^{-5}$\Msun, but much of the accretion occurs at rates large enough to trigger nuclear burning on the WD surface.  We discuss the consequences of our results for the observed outbursts of RS Oph, and speculate as to their likely origin.


\section{Model}
\subsection{Model}
The evolution of the gas surface density, $\Sigma(R,t)$ in the accretion disc around the primary is governed by the equation \citep{lbp74,pringle81}
\begin{equation}\label{eq:diff}\label{eq:gas_evo}
\frac{\partial \Sigma}{\partial t} = \frac{1}{R}\frac{\partial}{\partial R}\left[ 3R^{1/2} \frac{\partial}{\partial R}\left(\nu \Sigma R^{1/2}\right) - \frac{2 \Lambda \Sigma R^{3/2}}{(GM_1)^{1/2}}\right] + \dot{\Sigma}_{\mathrm {infall}}(R,t) \, .
\end{equation}
Here $t$ is time, $R$ is cylindrical radius, $\nu$ is the kinematic viscosity and $M_1$ is the mass of the primary.  The first term on the right-hand side represents viscous evolution of the disc, while the source term $\dot{\Sigma}_{\mathrm {infall}}(R,t)$ represents the mass infall on to the disc.  The second term describes the response of the disc to the tidal torque from the secondary.  $\Lambda(R,R_2)$  is the rate of specific angular momentum transfer from the secondary to the disc \citep[e.g.,][]{lp86}, given by
\begin{equation}
\Lambda(R,R_2) = - \frac{q^2 GM_1}{2R} \left(\frac{R}{\Delta_{\mathrm p}}\right)^4\end{equation}
where $q = M_2/M_1$ is the binary mass ratio and 
\begin{equation}
\Delta_{\mathrm p} = \textrm{max}(H,|R-R_2|) \, .
\end{equation}
Here $R_2$ is the binary separation and and $H = c_{\mathrm s}/\Omega$ is the disk scale-height.  We adopt a standard \citet{ss73} $\alpha$-viscosity 
\begin{equation}
\nu = \alpha  c_{\mathrm s}^2 / \Omega \, ,
\end{equation}
where $c_{\mathrm s}$ is the local sound speed, $\Omega = \sqrt{G M_1/R^3}$ is the Keplerian angular frequency, and the dimensionless parameter $\alpha$ specifies the efficiency of angular momentum transport \citep[thought to be due to the magnetorotational instability, e.g.,][]{bh91,bh98}.  We use a simplified form for the energy equation \citep[e.g.][]{cannizzo93}
\begin{equation}\label{eq:temp}
\frac{\partial T_{\mathrm c}}{\partial t} = \frac{2 (Q_+ - Q_-)}{c_{\mathrm P}\Sigma} - u_R
\frac{\partial T_{\mathrm c}}{\partial R} \, .
\end{equation}
Here $T_{\mathrm c}$ is the midplane temperature and $Q_+/Q_-$ are the instantaneous heating/cooling rates.  The radial velocity $u_R$ in the advective term in Equation \ref{eq:temp} is taken to be the vertically averaged value, and is computed as
\begin{equation}\label{eq:u_r}
u_R = - \frac{3}{\Sigma R^{1/2}} \frac{\partial}{\partial R} \left(\nu \Sigma R^{1/2}\right) \, .
\end{equation}
The specific heat capacity $c_{\mathrm P}$ is approximately constant for most temperatures of interest, but rises dramatically at $\simeq10^4$K (due to the ionization of atomic hydrogen).  We adopt the functional form for $c_{\mathrm P}$ from \citet{cannizzo93}.  The heating term $Q_+$ is evaluated as
\begin{equation}\label{eq:Q_plus}
Q_+ = Q_{\nu} + Q_{\mathrm {tid}} + Q_{\mathrm {hs}} \, .
\end{equation}
$Q_{\nu}$ is the heating from viscous dissipation, given by
\begin{equation}
Q_{\nu} = \frac{9}{8}\nu \Sigma \Omega^2 \, .
\end{equation} 
The heating due to the tidal torque from the secondary is evaluated as \citep[e.g.,][]{lodato09}
\begin{equation}
Q_{\mathrm {tid}} =  [\Omega_2 - \Omega(R)] \Lambda(R) \Sigma(R) dR \, ,
\end{equation}
where $\Omega_2$ is the orbital frequency of the binary.  The heat input at the ``hot-spot'' where the accretion stream lands on the disc is given by \citep{lasota01}
\begin{equation}\label{eq:hotspot}
Q_{\mathrm {hs}} = \eta \frac{G M_1 \dot{M}_{\mathrm {infall}}}{2 R_{\mathrm {out}}}\frac{1}{2 \pi \Delta R_{\mathrm {hs}}} \exp\left(\frac{R - R_{\mathrm {out}}}{ \Delta R_{\mathrm {hs}}} \right) \, .
\end{equation}
Here $\eta = 1$ is an (arbitrary) efficiency factor, $\dot{M}_{\mathrm {infall}}$ ($= \int 2 \pi R \dot{\Sigma}_{\mathrm {infall}} dR$) is the accretion rate from the secondary, $R_{\mathrm {out}}$ is the outer disc radius (computed as the radius where $\Sigma$ falls below 1 g cm$^{-2}$), and $\Delta R_{\mathrm {hs}} = 0.1 R_{\mathrm {out}}$ is the width of the heated region.  In practice viscous heating dominates: $Q_{\mathrm {hs}}$ is almost completely negligible, and tidal heating is only significant very close the the outer edge of the disc (and even then is not dominant).

We adopt a ``one-zone'' model for cooling \citep[as used by, for example,][]{jg03}, so that
\begin{equation}\label{eq:onezone}
Q_- = \frac{8}{3} \frac{\tau}{1 + \tau^2} \sigma_{\mathrm {SB}} \left(T_{\mathrm c}^4 - T_{\mathrm {min}}^4\right) \, .
\end{equation}
Here, the second term allows for a smooth transition between the limits of optically thick and optically thin cooling, $\sigma_{\mathrm {SB}}$ is the Stefan-Boltzmann constant, and the last term is prescribed in a manner than enforces a minimum disc temperatute $T_{\mathrm {min}} = 10$K.  The vertical optical depth $\tau$ is computed as
\begin{equation}
\tau = \frac{1}{2}\kappa(T_{\mathrm c},\rho_{\mathrm c}) \Sigma \, ,
\end{equation}
with the midplane density $\rho_c$ evaluated from the surface density by assuming that the disc is vertically isothermal:
\begin{equation}
\rho_{\mathrm c} = \frac{\Sigma}{\sqrt{2\pi} (c_{\mathrm s}/\Omega)} \, .
\end{equation}
The local sound speed is essentially a means of keeping track of the local pressure, and includes contributions from both gas and radiation pressure:
\begin{equation}\label{eq:c_s}
c_s^2 = \frac{{\cal R} T_c}{\mu} + \frac{4 \sigma_{{\mathrm {SB}}} T_c^4}{3 c \rho_c}
\end{equation}
Here $\cal R$ is the gas constant, $\mu$ is the mean molecular weight, and $c$ is the speed of light.  In practice, radiation pressure is only significant at high temperatures.  We adopt standard numerical fits for both $\kappa(T,\rho)$ and $\mu(T,\rho)$: that of \citet{zhu07,zhu08} for the opacity, and that of \citet{hure00} for the mean molecular weight.

\subsubsection{Numerical Method}
We solve Equations \ref{eq:diff} and \ref{eq:temp} using an explicit finite-difference method, on a grid that is equispaced in $R^{1/2}$ \citep*[e.g.,][]{pvw86}, using 1587 grid cells spanning the range $[10^{10}\mathrm {cm},10^{13}\mathrm {cm}]$.  We use a staggered grid, with scalar quantities evaluated at at cell centres and vectors (velocities) at cell edges.  The derivative in the advective term (Equation \ref{eq:u_r}) is evaluated as the first-order, up-winded value.  We adopt zero-torque boundary conditions at both the inner and outer boundaries, by setting $\Sigma=0$ in the boundary cells.

For our fiducial model we adopt parameters consistent with those observed for the RS Oph system \citep[e.g.,][]{hk00,schaefer09}: $M_1 = 1.35$\Msun, $q = M_2/M_1 = 0.6$, $R_2 = 2.1\times10^{13}$cm, and $P = 2\pi/\Omega_2 = 453.5$ days.  (Note that these parameters are only self-consistent at the 10\% level, due to observational uncertainties and rounding errors.)  We assign a small initial surface density of $\Sigma = 1$g cm$^{-2}$ at all radii, and allow the disc to be built-up self-consistently as material accretes from the secondary.

\subsection{S-curves}
It is well known that normal WD accretion discs, such as those around CVs, are thermally unstable at $T_{\mathrm c} \sim 10^4$K \citep[see, e.g.,][and references therein]{lasota01}, due to the rapid increase in opacity with temperature as hydrogen is ionized.  However, it is also well-known that thermal instability alone cannot reproduce the behaviour of CVs; instead we require the thermal--viscous instability, where the efficiency of angular momentum transport also increases substantially as hydrogen is ionized.  This behaviour is incorporated in our model by assuming that $\alpha$ varies with temperature as follows:
\begin{equation}\label{eq:alpha}
\alpha(T_{\mathrm c}) = \frac{\alpha_{\mathrm {hot}} - \alpha_{\mathrm {cold}}}{1 + \exp\left(\frac{T_{\mathrm {crit}} - T_{\mathrm c}}{\Delta T}\right)} + \alpha_{\mathrm {cold}} \, .
\end{equation}
This form allows for a smooth transition between the ``cold-state'' value $\alpha_{\mathrm {cold}}$ and the ``hot-state'' value $\alpha_{\mathrm {hot}}$, as the temperature $T$ rises above some critical value $T_{\mathrm {crit}}$.  In our fiducial model we set $T_{\mathrm {crit}} = 5000$K and $\Delta T = 250$K, with $\alpha_{\mathrm {cold}} = 0.01$ and $\alpha_{\mathrm {hot}} = 0.1$.

\begin{figure}
\centering
       \resizebox{\hsize}{!}{
       \includegraphics[angle=270]{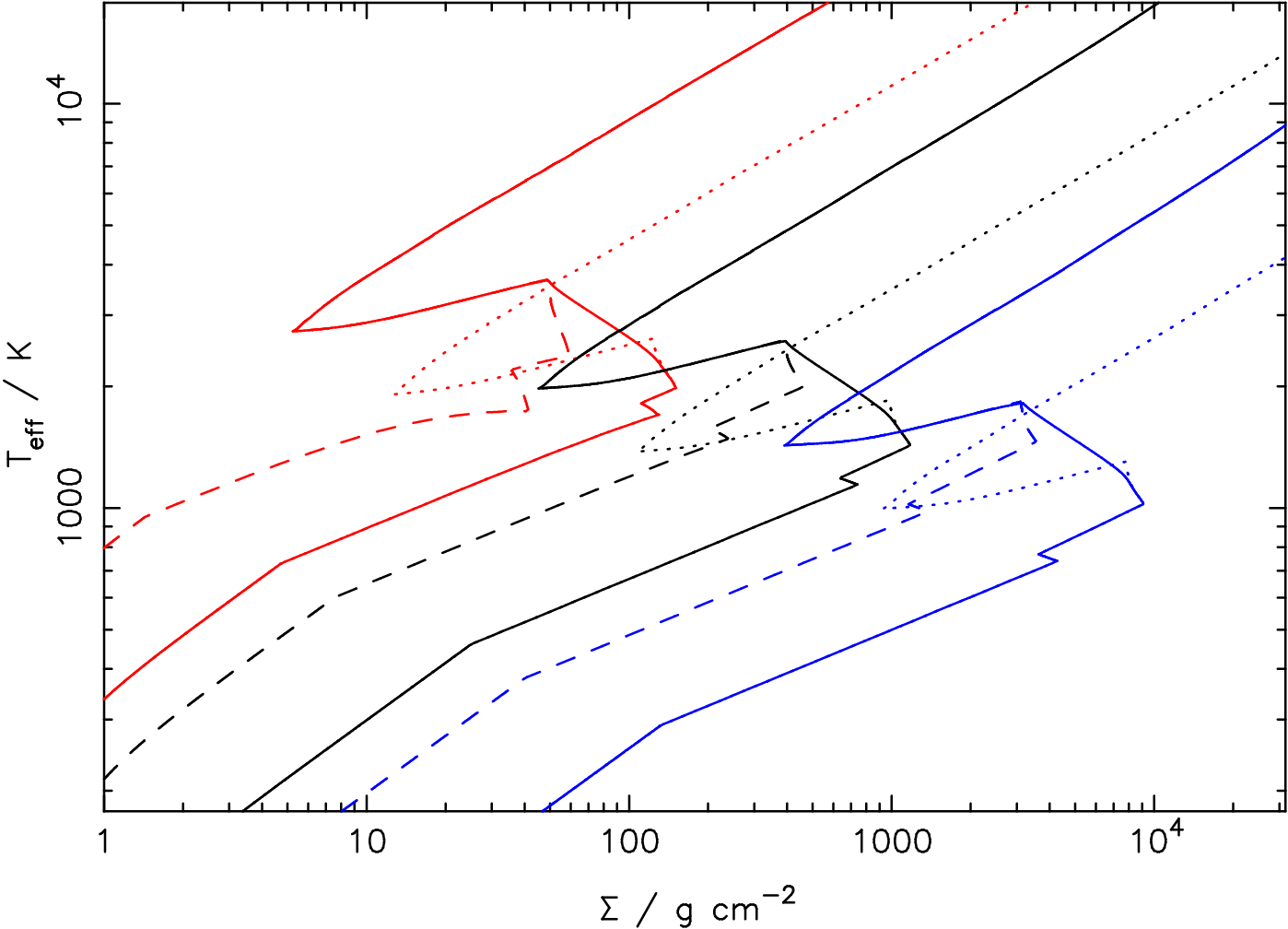}
       }
       
       \vspace*{6pt}
       
       \resizebox{\hsize}{!}{
       \includegraphics[angle=270]{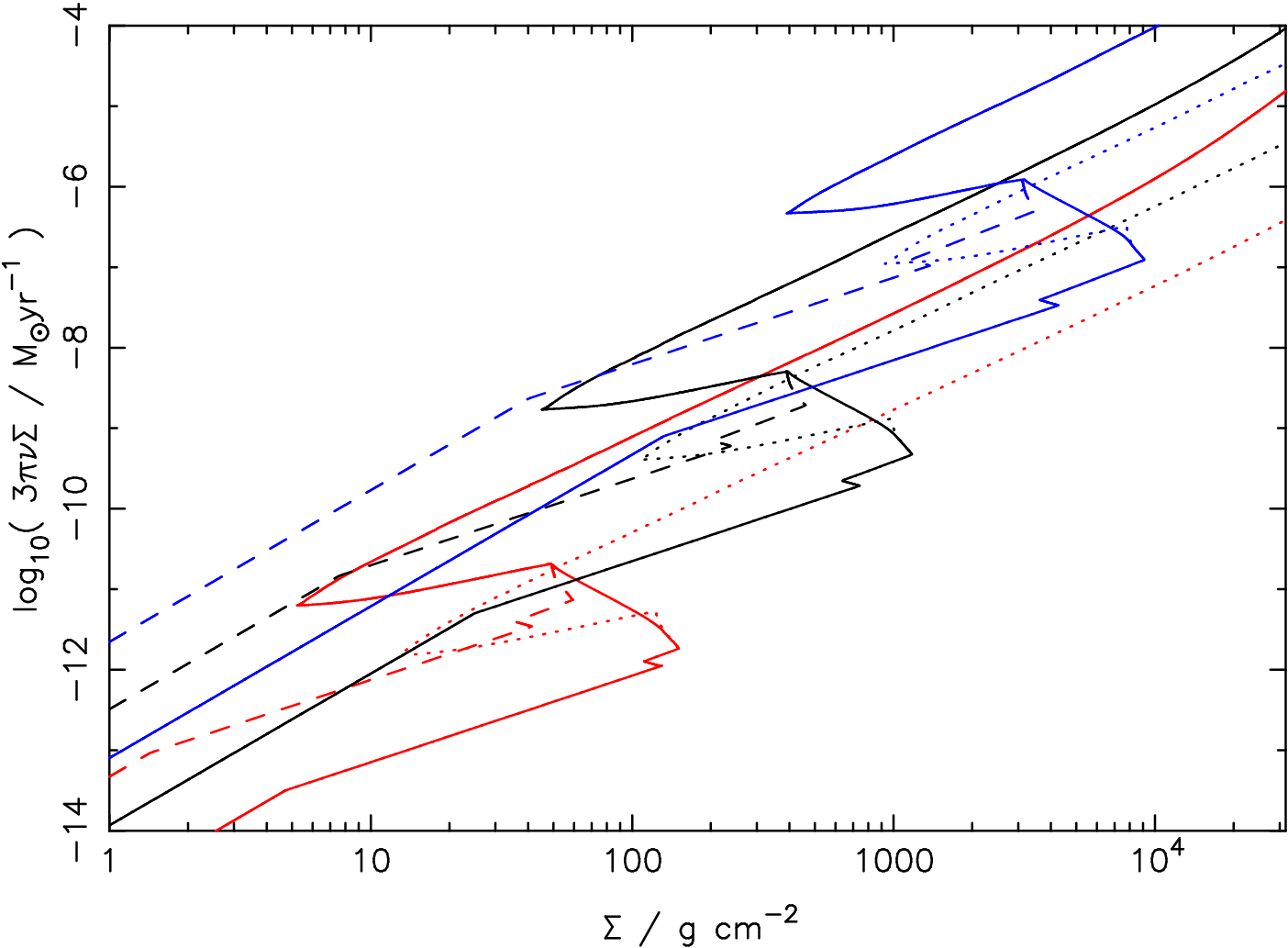}
       }
       \caption{S-curves for our disc model, computed for the parameters of RS Oph system.  The upper panel shows the curves in the standard $\Sigma$--$T_{\mathrm {eff}}$ plane; the lower panel the more instructive $\Sigma$-$\dot{M}$ plane.  In each panel the red, black and blue curves represent $R = 10^{10}$, $10^{11}$ \& $10^{12}$cm respectively.  The dotted and dashed curves are computed for fixed $\alpha$ (cold and hot, respectively), while the solid curve uses the temperature-dependent $\alpha$-prescription described in Equation \ref{eq:alpha} (which ``jumps'' from $\alpha_{\mathrm {cold}}$ on the lower stable branch to $\alpha_{\mathrm {hot}}$ on the upper stable branch).}     
           \label{fig:S_curves}
\end{figure}

From these equations we can determine the range of parameter space where the disc will be unstable by assuming local thermodynamic equilbrium ($Q_{\nu} = Q_-$).  At a given radius in the disc we can then solve for $\Sigma$ as a function of $T_c$, using Equations \ref{eq:Q_plus}--\ref{eq:c_s} above.  As the discs are optically thick everywhere we can define the effective (or photospheric) temperature $T_{\mathrm {eff}} = T_{\mathrm c} \tau^{-1/4}$, and derive the ``S-curves'' characteristic of disc instability models.  These are shown in Fig.\,\ref{fig:S_curves}.  

The S-curves from our simple, one-zone model are somewhat crude, and lack the detail obtainable with more sophisticated models \citep[e.g.,][]{hameury98}, but they capture the essential ingredients of the model well.  We immediately see from Fig.\,\ref{fig:S_curves} that a sufficiently large disc is unstable over a very wide range in accretion rates: if the disc is $\sim10^{12}$cm in size, then it will be unstable at some radius for all accretion rates in the range $10^{-12}$\Msunyr $\lesssim \dot{M} \lesssim 10^{-5}$\Msunyr.  Further tests indicate that the shapes of the S-curves vary only slightly with changes the numerical parameters $T_{\mathrm {crit}}$ or $\Delta T$, and that the general behaviour is robust.

\subsection{Emitted spectrum model}
In order to compare to observations, we compute the (continuum) emission from the binary system at each time-step.  The model has three contributions: the emission from the disc, the emission from the boundary layer on the WD surface, and the emission from the secondary.  The secondary is assumed to be a Roche-lobe filling giant (with photospheric radius 0.7 times the Roche radius, $R_* \simeq 120$R$_{\odot}$), and is modelled as black-body with temperature $T_*$.  The emission from the disc is computed by assuming that each annulus emits as a black body with temperature $T_{\mathrm {eff}}$.  To compute the emission from the primary we assume that half of the accretion luminosity is radiated in a boundary layer on the WD surface.  If the boundary layer radiates as a black-body, we therefore have
\begin{equation}\label{eq:BL}
4 \pi R_{\mathrm {wd}}^2 \sigma_{\mathrm {SB}} T_{\mathrm {bl}}^4 = \frac{G M_1 \dot{M}}{2 R_{\mathrm {wd}}} \, ,
\end{equation}
where the accretion rate $\dot{M}$ is computed at the inner boundary.  If the WD radius $R_{\mathrm {wd}}$ is known Equation \ref{eq:BL} uniquely specifies the boundary layer temperature $T_{\mathrm {bl}}$.  We adopt a radius $R_{\mathrm {wd}} = 2.5\times10^{8}$cm \citep[consistent with a WD mass of 1.35\Msun, e.g.,][]{althaus05} and a secondary surface temperature of $T_*=3500$K \citep{schaefer09}.  With this parametrization, the total accretion luminosity scales as
\begin{eqnarray}
\nonumber 
L_{\mathrm {bl}} = 1.1\times10^{37} \, {\mathrm {erg}} \, \mathrm s^{-1} \, \left(\frac{M_1}{1.35\mathrm M_{\odot}}\right) \left(\frac{\dot{M}}{5\times10^{-7}\mathrm M_{\odot}\mathrm {yr}^{-1}}\right) \\ \times \left(\frac{R_{\mathrm {wd}}}{2.5\times10^8\mathrm {cm}}\right)^{-1} \, ,
\label{eq:L_bl}
\end{eqnarray}
and the boundary layer temperature as
\begin{eqnarray}
\nonumber
T_{\mathrm {bl}} = 7.1 \times 10^5 \, \mathrm K  \, \left(\frac{M_1}{1.35\mathrm M_{\odot}}\right)^{1/4} \left(\frac{\dot{M}}{5\times10^{-7}\mathrm M_{\odot}\mathrm {yr}^{-1}}\right)^{1/4} \\  \times \left(\frac{R_{\mathrm {wd}}}{2.5\times10^8\mathrm {cm}}\right)^{-3/4} \, .
\label{eq:T_bl}
\end{eqnarray}
The luminosity of the secondary is $L_* = 4\pi R_*^2 \sigma_{\mathrm {SB}} T_*^4 = 7.5\times10^{36}$erg\,s$^{-1}$.

The inner edge of our computational grid is at $R = 10^{10}$cm, and is thus an order of magnitude larger than the radius of the WD.  It is very computationally expensive to extend the grid to smaller radii, because the innermost grid cells require very short time-steps during outburst.  Neglecting this inner part of the disc has a negligible effect on the models, as the boundary plays no significant role in the evolution, but it can have a potentially important effect on the SED (as we neglect the hottest part of the disc).  In outburst the disc follows an almost perfect $T_{\mathrm {eff}} \propto R^{-3/4}$ power-law, so we also compute an additional SED in which we extrapolate the disc temperature beyond the inner edge of our computational domain.  We extrapolate in to $R = 8\times10^8$cm, using a $T_{\mathrm {eff}} \propto R^{-3/4}$ power-law and normalising at $R = 5 \times 10^{10}$cm.  This extrapolation is somewhat dubious, as departure from a simple power-law is likely in the boundary layer, but as it has only a small effect on the resulting light-curves (see Fig.\,\ref{fig:V_band}) we do not consider it to be a significant source of uncertainty.

 
\section{Results}\label{sec:res}
The model described above essentially has two free parameters: the rate of accretion from the secondary, and the angular momentum of the accreted material.  These are parametrized in our models as the integrated accretion rate and the radial profile of the source term $\dot{\Sigma}_{\mathrm {infall}}$.  For simplicity, we assume that the profile is a $\delta$-function at the circularisation radius $R_{\mathrm c}$, and thus the model is completely specified by the two parameters $\dot{M}_{\mathrm {infall}}$ and $R_{\mathrm c}$.

\begin{figure}
\centering
       \resizebox{\hsize}{!}{
       \includegraphics[angle=270]{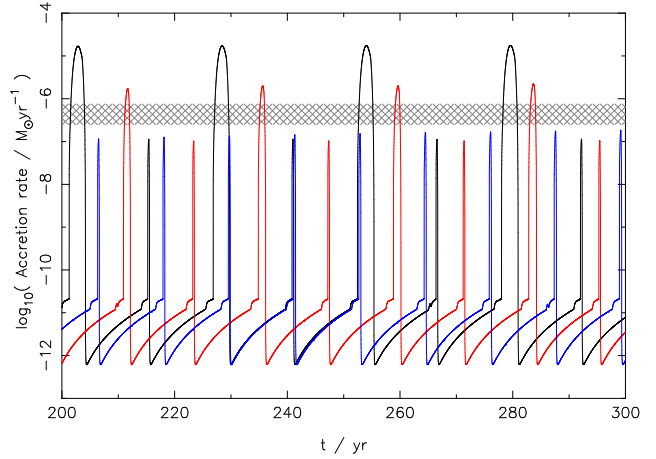}
       }
       \caption{Accretion rates on to the WD as a function of time for the fiducial models (computed at the inner boundary of the grid).  The black, red and blue curves represent the results for $\log_{10} (\dot{M}_{\mathrm {infall}}/$\Msunyr)$ = -6.0$, $-7.0$ \& $-8.0$ respectively.  The offsets between the outburst times for the different models is an artefact of our initial conditions, as the models with lower infall rates take longer to build up sufficient mass to trigger their initial outburst.  Once the limit cycle has been established all of the models show a recurrence time-scale of $\simeq 12$yr.  Note also the pattern of alternating large and small outbursts in the models with larger infall rates.  The hatched region shows the approximate locus of the stable nuclear burning band for a 1.35\Msun WD, based on the calculations of \citet{sb07}.}     
           \label{fig:mdot_time}
\end{figure}

We assume that accretion from the secondary proceeds via Roche lobe overflow.  In this case the circularisation radius can be calculated explicitly from the binary parameters \citep[e.g.,][]{wynn08}, and we find that $R_{\mathrm c} = 1.74\times10^{12}$cm.  We compute the evolution of models for a range of infall rates $\log_{10} (\dot{M}_{\mathrm {infall}}/$\Msunyr)$ = -8.0$, $-7.5$, $-7.0$, $-6.5$, $-6.0$.  In each case the disc builds up due to the accretion of mass from the secondary, and once the disc is sufficiently massive we see repeated outbursts due to the thermal--viscous instability in the disc.  The outbursts are of the ``inside-out'' type, and are triggered close to the inner boundary.  After some initial transients have died out a stable limit cycle is established, with the all of the discs showing repeated outbursts with a recurrence time-scale of $\simeq 12$yr.  Fig.\,\ref{fig:mdot_time} shows the accretion rates on to the WD as a function of time.  The magnitude of the outbursts depends on how much mass has accumulated in the outer disc during quiescence, with higher values of $\dot{M}_{\mathrm {infall}}$ resulting in much larger outbursts.  The magnitude of any given outburst is determined by how far outward the ionization front can propagate, which in turn depends on the distribution of mass $\Sigma(R)$ at the moment the outburst is triggered.  For high values of $\dot{M}_{\mathrm {infall}}$ we see a pattern of alternating large and small outbursts, with the larger outbursts propagating much further into the outer disc and accreting substantially more mass.  Note, however, that in all cases the time-averaged accretion rate on to the WD is exactly equal to the rate of infall on to the disc from the secondary.

\begin{figure}
\centering
       \resizebox{\hsize}{!}{
       \includegraphics[angle=270]{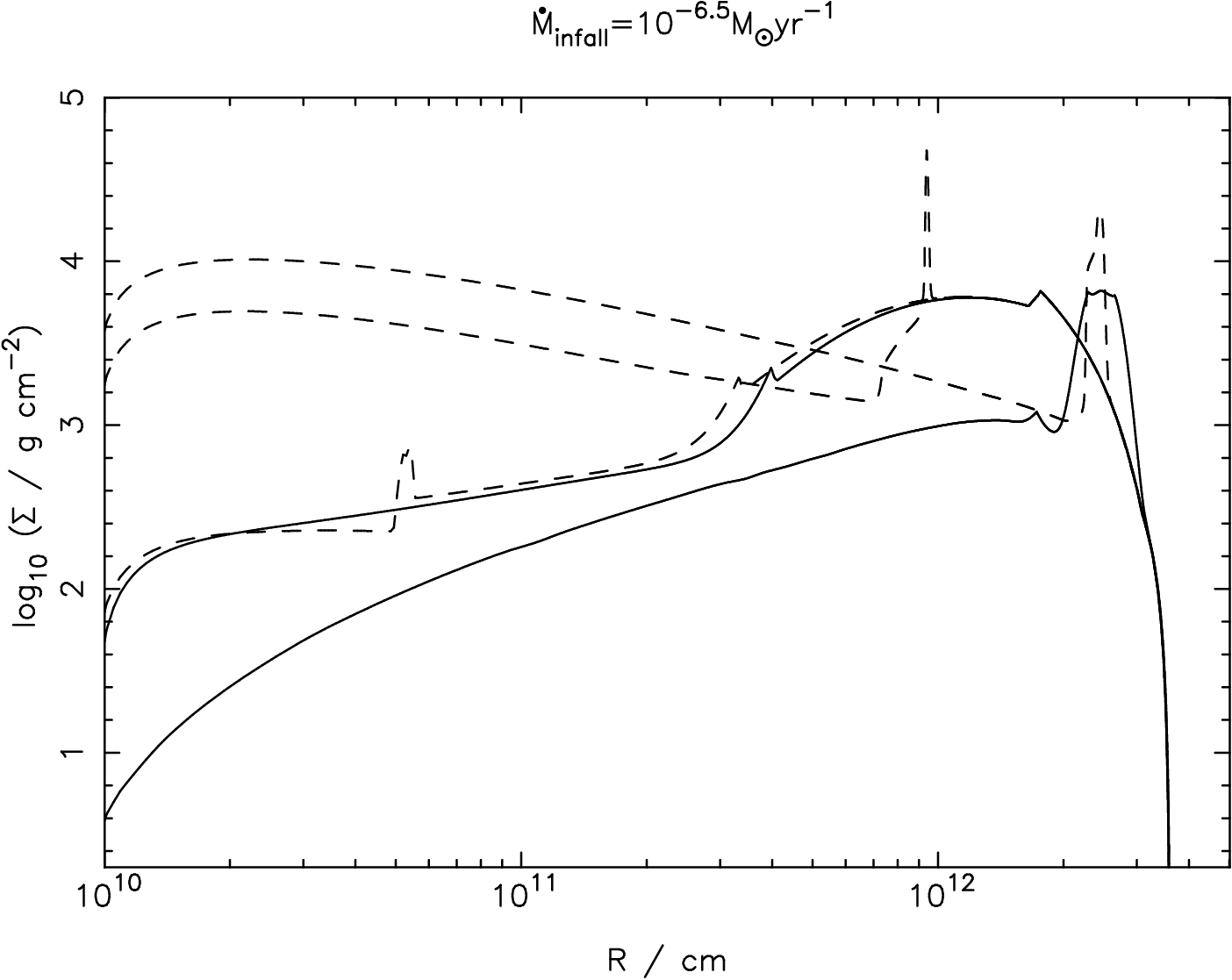}
       }
       
       \vspace*{12pt}
       
       \resizebox{\hsize}{!}{       
       \includegraphics[angle=270]{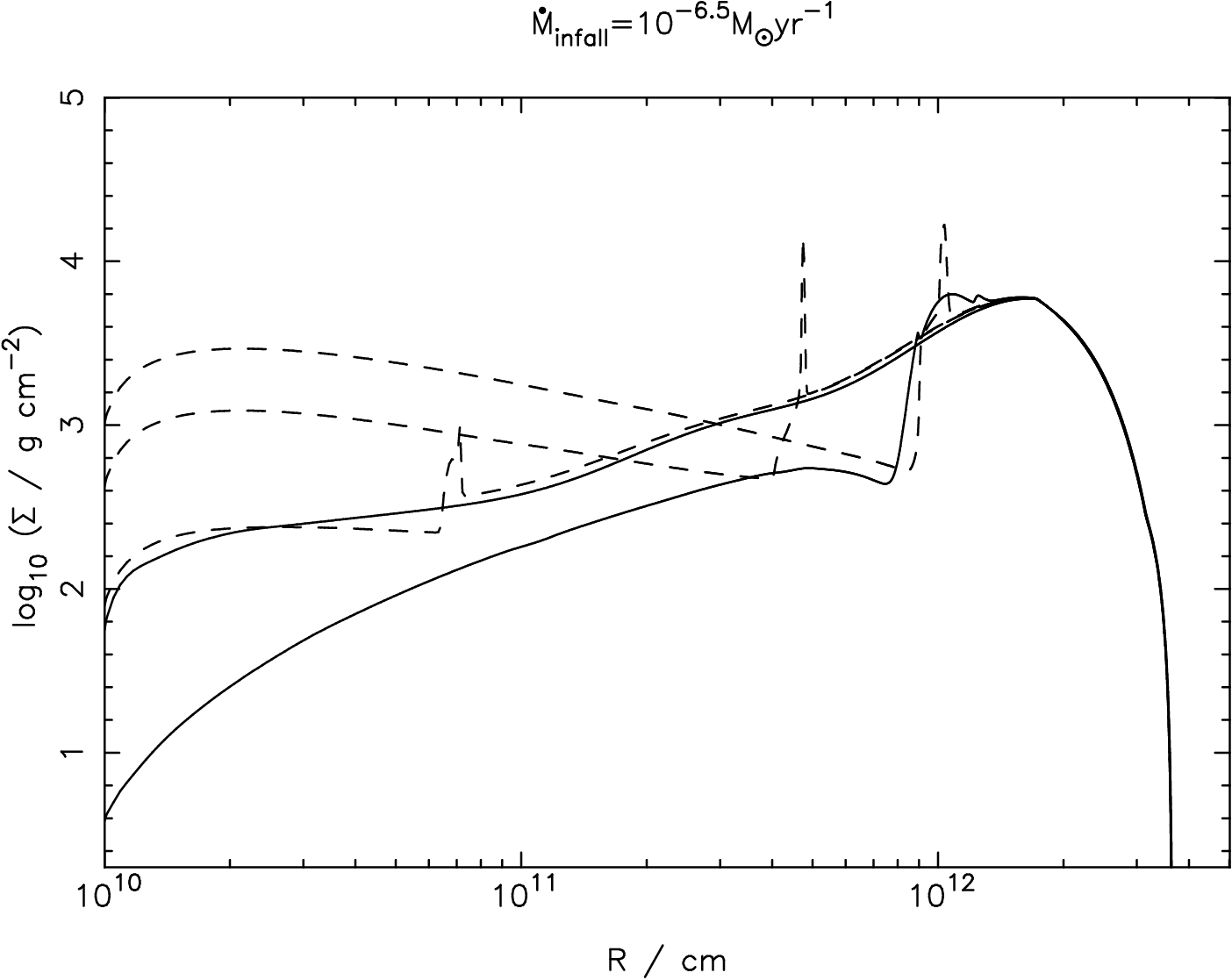}
       }
       \caption{Surface density evolution during an outburst in the model with $\log_{10} (\dot{M}_{\mathrm {infall}}/$\Msunyr)$ = -6.5$, for both large (top) and small (bottom) outbursts.  The upper and lower solid curves respectively show the surface density profiles immediately before and after the outbursts; the dotted curves show evolution during the outburst.  The outbursts are triggered at small radii, and the ionization front then rapidly propagates outward through the disc.  In the upper panel the curves are plotted 1.3, 1.7, 2.6 \& 3.55yr after the initial profile; in the lower panel the corresponding times are 1.3, 1.5, 1.95 \& 2.45yr.}      
	\label{fig:sigma_time}
\end{figure}

The fact that the recurrence time-scale is independent of $\dot{M}_{\mathrm {infall}}$ can be readily understood.  Fig.\,\ref{fig:sigma_time} shows the surface density evolution of the disc during both large and small outbursts.  Following an outburst the disc is essentially empty at small radii, but the structure at larger radii depends on how far the outburst was able to propagate into the disc.  This in turn depends on the size of the mass reservoir that has built up in the outer disc (at radii $\sim R_{\mathrm c}$): if the outer disc is sufficiently massive then the ionization front can propagate to larger radii.  We see from Fig.\,\ref{fig:S_curves} that the disc is expected to be unstable at some radius for all of the infall rates considered here, so the recurrence time-scale is simply the time required for the surface density at some radius to exceed the critical value for instability.  This first happens at small radii, typically a few times $10^{10}$cm (hence the ``inside-out'' outbursts), so the recurrence time-scale is essentially the time required for gas to accrete from $R_{\mathrm c}$ down to the triggering radius.  This in turn is simply the viscous time-scale at $R_{\mathrm c}$, which in the cold state depends primarily on $\alpha_{\mathrm {cold}}$ (with only a very weak dependence on $\Sigma$, through the dependence of the viscous time-scale on the disc temperature).  Larger infall rates therefore lead to larger and longer-duration outbursts, but do not alter the recurrence time-scale significantly.  This is verified by computing additional models with $\alpha_{\mathrm {cold}} = 0.005$: here, the recurrence time-scale is consistently $\simeq 25$yr, almost exactly a factor of two longer than in the models with $\alpha_{\mathrm {cold}} = 0.01$.

\begin{figure}
\centering
       \resizebox{\hsize}{!}{
       \includegraphics[angle=270]{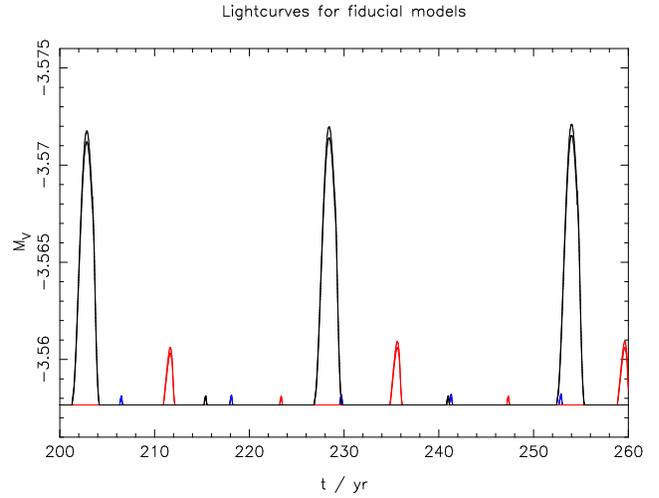}
       }
       \caption{Simulated $V$-band lightcurves for the fiducial models (in absolute AB-magnitudes): note the small dynamic range on the vertical axis.  As before, the black, red and blue curves are for $\log_{10} (\dot{M}_{\mathrm {infall}}/$\Msunyr$) = -6.0$, $-7.0$ \& $-8.0$ respectively.  In each case the thinner line shows the light-curve if the disc is extrapolated beyond the inner edge of the grid (as described in the text); in practice, we see that this makes a negligible difference to the optical light-curve.  The largest outbursts only marginally exceed the quiescent flux level, and are barely detectable above the emission from the giant secondary.  By contrast, during outbursts RS Oph is observed to brighten by 6--7\,mag in the $V$-band \citep{schaefer10}.}     
           \label{fig:V_band}
\end{figure}

Fig.\,\ref{fig:V_band} shows the simulated $V$-band light-curves for the fiducial models.  Even in the largest outbursts the luminosity of the disc only barely exceeds that of the giant, and at optical wavelengths the net effect is a brightening only at the $\sim 0.01$\,mag level.  By contrast, in the $V$-band RS Oph is observed to brighten by a factor of 100--1000 (6--7\,mag) during its RN events \citep[e.g.,][]{schaefer10}\footnote{The bolometric luminosity of the outbursts remains poorly constrained, primarily due to the large uncertainty in the distance to RS Oph, but is estimated to peak at $\sim10^{38}$erg\,s$^{-1}$ \citep[e.g.,][]{schaefer09,osborne11}.}.  The difference between the ``extrapolated'' and standard light-curves from our models is essentially negligible, as the inner disc (the extrapolated part) contributes negligibly to the $V$-band flux.  In the largest outbursts the bolometric luminosity of the boundary layer (effectively the accretion luminosity) exceeds that of the secondary by almost two orders of magnitude, as can be seen from Equation \ref{eq:L_bl}.  However, because of the high temperature of the boundary layer (Equation \ref{eq:T_bl}) most of this luminosity is emitted at short wavelengths, and at optical wavelengths the outbursts are only marginally brighter than the secondary.

\begin{figure}
\centering
       \resizebox{\hsize}{!}{
       \includegraphics[angle=270]{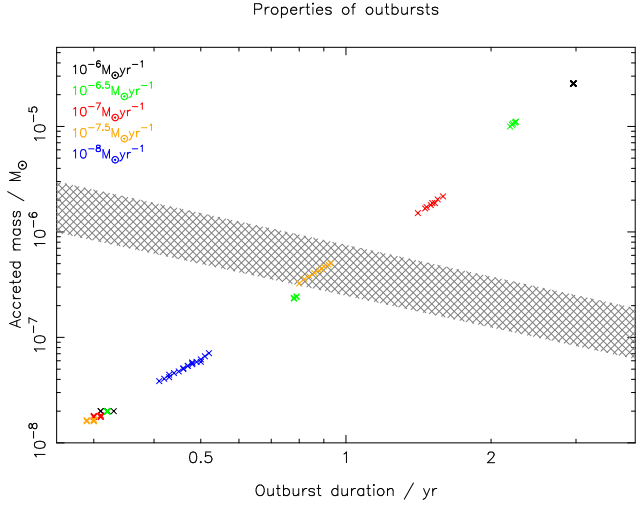}
       }
       \caption{Duration and total accreted mass for 17 consecutive outbursts in each of the fiducial models.  The smaller outbursts are comparable for all values of $\dot{M}_{\mathrm {infall}}$, but the larger outbursts increase sharply in magnitude as $\dot{M}_{\mathrm {infall}}$ increases.  The largest outbursts result in the accretion of $>10^{-5}$\Msun on to the WD on time-scales of 2--3yr.  The hatched region denotes where the mean accretion rate during an outburst lies in the stable nuclear burning band for a 1.35\Msun WD, based on the calculations of \citet{sb07}. }     
           \label{fig:outbursts}
\end{figure}

During outbursts, the observed optical lightcurves of RS Oph show a rapid rise (on a time-scale of days) followed by decay back to the quiescent level on a time-scale $\sim 100$ days \citep[e.g.,][]{schaefer10}.  The total duration of the outbursts in our models is somewhat longer than this, by a factor of a few (see Fig.\,\ref{fig:outbursts}), but given the uncertainties in our model parameters (notably $\alpha_{\mathrm {hot}}$) is not clear whether this discrepancy is significant.  However, the shape of the lightcurves in our model (which are essentially symmetric about the outburst peaks) is inconsistent with the observed lightcurves.  Our model therefore fails to reproduce both the magnitude and time evolution of the outbursts that are observed at optical wavelengths, so disc instability alone cannot explain the outbursts seen in RS Oph.

Fig.\,\ref{fig:outbursts} shows the properties of the outbursts for the fiducial model set.  For small values of $\dot{M}_{\mathrm {infall}}$ the outbursts are short in duration, and the WD accretes $<10^{-7}$\Msun.  However, the magnitude of the outbursts increases sharply with increasing $\dot{M}_{\mathrm {infall}}$, and for values $\dot{M}_{\mathrm {infall}} \gtrsim 3 \times 10^{-7}$\Msunyr the total mass accreted during an outbursts can exceed $10^{-5}$\Msun.  Also shown in Fig.\,\ref{fig:outbursts} (see also Fig.\,\ref{fig:mdot_time}) is the approximate location of the stable nuclear burning band, based on the calculations of \citet{sb07}.  Only for accretion rates within this region accreted hydrogen can burn stably on the WD surface: for lower accretion rates the WD simply accretes unprocessed gas from the disc, while for higher accretion rates the WD surface becomes thermally unstable.  We see that in our model the accretion rate on to the WD during outburst can span all three of these regimes, depending on the accretion rate from the secondary on to the disc.  

Given the similarity between the recurrence time-scale of the disc instability in our model and that of the observed outbursts, it is tempting to speculate that accretion due to the disc outbursts could be responsible for triggering a thermonuclear explosion on the disc surface.  However, this requires the WD to accrete unprocessed hydrogen from the disc, while in most of our models the accretion rate exceeds that at which the accreted material will undergo nuclear burning as soon as it reaches the WD surface.  Indeed, in all our models the mass of unprocessed material accreted by the WD (i.e., mass accreted below the nuclear burning band) is $<10^{-7}$\Msun~per outburst.  This suggests that the time-scale required to accrete enough unprocessed hydrogen to trigger a thermonuclear runaway is $\gtrsim 200$yr.  This is much longer than the observed recurrence time-scale, and argues strongly against thermonuclear runaway as the origin of the observed outbursts.

By contrast, Fig.\,\ref{fig:mdot_time} shows that in most cases the accretion rate during outbursts is sufficient to allow stable nuclear burning on the WD surface.  For a 1.35\Msun WD the efficiency of nuclear burning exceeds that of accretion by a factor of $\sim 20$ \citep{frank02}, so comparison with Equation \ref{eq:L_bl} shows that stable nuclear burning can yield a luminosity\footnote{Alternatively one can estimate this value as $\epsilon\dot M_{\mathrm {steady}}c^2 \simeq 1$--$3\times10^{38}$erg\,s$^{-1}$, where $\epsilon = 0.007$ is the efficiency of hydrogen burning.} of $\sim 10^{38}$erg\,s$^{-1}$.  This is comparable to the observed outburst luminosities, and shows that stable nuclear burning is an energetically plausible mechanism for powering the outbursts.  However, in most outbursts the accretion rate exceeds the upper envelope of the stable nuclear burning band, so it is not obvious that stable nuclear burning can be sustained. What happens in this case is far from clear.  If the accretion rate exceeds the stable burning limit the accretion flow becomes thermally unstable and expands, in a manner similar to a red giant atmosphere, but the time-scale for this expansion depends sensitively on several unknown parameters.  It is possible that such an event could be responsible for the observed outbursts of RS Oph \citep[see also][]{kp09}, but detailed modelling of this process is beyond the scope of this paper.  We note also that this type of mechanism, where disc instability acts as the trigger for the observed outbursts, may provide an explanation for the recently-detected pre-outburst signal in the lightcurve of RS Oph \citep{adamakis11}.


\section{Discussion}\label{sec:dis}
Our simple one-dimensional disc model has several limitations, the effects of which may be important in certain circumstances.  The most significant of these is our almost total neglect of the disc's vertical structure, the so-called ``one-zone'' approximation (Equation \ref{eq:onezone}).  In general this is a reasonable approximation to make in thin discs, but it can break down in regions where the opacity varies significantly over the vertical extent of the disc.  This can occur during disc instability outbursts (as the instability is triggered by the steep dependence of opacity with temperature), and an improved treatment would consider the vertical structure of the disc in detail \citep[e.g.,][]{bl94,hameury98}.  In practice, however, this procedure can be well approximated in one-dimensional models by modifying the resulting S-curves appropriately, and the results from such models are broadly consistent with our results.  Thus while the precise details of our model thus depend on the shape of our derived S-curves [which in turn depend on the chosen functions $\kappa(\rho,T)$ and $\mu(\rho,T)$], the qualitative behaviour should be robust.

In addition our one-dimensional treatment explicitly neglects non-axisymmetric effects, in particular the tidal field of the binary and the localised nature of the accretion hot-spot.  Our azimuthally-averaged approximation of the hot-spot heating (Equation \ref{eq:hotspot}) finds that the magnitude of the bulk heating is negligible, so while the hot-spot may have observational signatures the treatment of the hot-spot is unlikely to affect the disc evolution significantly.  Further, because the disc instability outbursts are triggered at small radii, tidal effects (which primarily affect the disc near its outer edge) will have a negligible influence on the disc's evolution.


\subsection{Burning during outbursts?}
We have shown that the accretion disc in RS Oph is likely to undergo dwarf nova outbursts, with a characteristic recurrence time of order $\simeq 12$~yr. There is a pattern of alternating large and small outbursts, with the large outbursts recurring at intervals of $\simeq 25$yr, quite similar to the observed recurrence time-scale of RS Oph's outbursts.  We also find that even during the largest outbursts the predicted accretion luminosity is insufficient to explain the observed optical outbursts of RS Oph.  However, in all cases the accretion rate during these large outbursts crosses the band for steady nuclear burning, and this offers a plausible explanation of the observed outbursts.  Is it not clear how the system responds to both surface burning and a rapidly changing accretion rate, but our result suggests that quasi-steady nuclear burning does occur at some point during these outbursts.  Within the model used in this paper it is difficult to go further, as the development of the large outbursts must be affected by the the high nuclear luminosity of the WD.  The effect is probably similar to soft X--ray transients, where X--ray irradiation by the central accreting neutron star or black hole traps the disc in the hot state and prolongs the outburst until the central regions of the disc are largely drained of matter \citep[e.g.,][]{kr98}.  The possibility that long-period dwarf novae outbursts might trigger steady nuclear burning was considered by \citet*{king03}, who showed that this may be a promising candidate for the progenitors of SNe Ia.  If stable nuclear burning can be sustained the WD mass will grow, and our model suggests that in this case the WD will be able to accrete $\gtrsim 10^{-6}$\Msun~per outburst (see Fig.\,5).  $10^4$--$10^5$ outbursts would therefore sufficient for the WD mass to increase by 0.01--0.1\Msun, which suggests that if RS Oph is the progenitor of a Type Ia SN it could reach $M_{\mathrm {Ch}}$ in $\lesssim1$Myr.

This scenario, however, raises a number of questions about the long-term evolution of the system.  If the WD mass is very close to $M_{\mathrm {Ch}}$ then a relatively small number of outbursts may be sufficient to trigger a SN.  The WD mass has not been measured directly, however, and if (as we argue) the RNe outbursts of RS Oph are not due to thermonuclear runaways there then the WD mass may be significantly lower, $\simeq 1.0$--1.3\Msun\footnote{It is also worth noting that if the system is to become a SN Ia then the initial (carbon-oxygen) WD mass must have been $\lesssim 1.2$\Msun~\citep[e.g.,][]{it84,nomoto84}.  A larger initial mass results in a oxygen-neon-magnesium WD, which will collapse into a neutron star rather than exploding as a SN Ia \citep[e.g.,][]{sn85}.}.  In this case we require the outburst cycle to be maintained for $\sim$1Myr in order to reach $M_{\mathrm {Ch}}$, which in turn requires that the disc instability limit-cycle be maintained over a large number of outbursts.  If a significant fraction of the WD mass is transferred from the secondary to the primary the binary will expand, and over time this may stop accretion via Roche lobe overflow.  However, this so-called nova hibernation is usually short-lived, and typically evolution of the secondary will re-start mass transfer relatively quickly \citep[e.g.,][]{shara86}.  The long-term stability of the outburst cycle is certainly an issue that must be considered, but as long as the WD is significantly more massive than the secondary there is no obvious reason to expect binary evolution to prevent the WD accumulating mass on secular time-scales.

In addition, repeated bursts of accretion in the stable nuclear burning band presumably result in a build-up of helium-rich material on the WD surface, and this could eventually lead to a helium nova explosion \citep{it91}.  The accretion scenario considered here (burning on arrival) is rather different from that typically used in helium nova models, and it is not clear whether the conditions for helium ignition will be met before the WD is able to go supernova.  However, detailed models of helium flashes on WDs find that helium novae eject significantly less mass than they accumulate prior to the explosion \citep[e.g.,][]{kh99}.  This is in contrast to conventional hydrogen novae, which typically eject all of the accreted mass, and suggests that the WD mass increases with time even if helium novae are triggered.  Further study of this issue is still needed, but growth of the WD mass towards $M_{\mathrm {Ch}}$ through repeated disc instability outbursts seems to be a plausible path to Type Ia SNe.


\section{Summary}
In this paper we have presented a one-dimensional time-dependent model for disc accretion in the RS Oph system.  We find that the large disc around the white dwarf primary is always unstable to the thermal--viscous instability, and we are able to study the properties of these outbursts in detail.  We find that:
\begin{itemize}
\item Disc instability in the RS Oph system naturally leads to repeated outbursts of accretion on $\sim 10$--20yr time-scales, similar to the recurrence time-scale of the observed outbursts.
\item The disc luminosity during outburst is not sufficient to explain the observed RNe outbursts.
\item The mass of unprocessed hydrogen accreted during the disc instability outbursts is small, $<10^{-7}$\Msun~per outburst.  This argues strongly against thermonuclear runaway as the origin of the observed outbursts.
\item It seems possible that disc instability combined with surface nuclear burning can explain the observed RNe outbursts.  The recurrence time-scale is naturally consistent with disc instability, and the luminosity of the observed outbursts is comparable to that of steady nuclear burning.
\item In our models the majority of outbursts attain (or exceed) the accretion rate required for steady nuclear burning.  This suggests that the white dwarf is thus able to grow to the Chandrasekhar mass on $<$Myr time-scales, and we suggest that RS Oph and similar systems may therefore be possible progenitors for Type Ia supernovae.
\end{itemize}

\section*{Acknowledgements}
We are grateful to Kim Page for useful discussions, and thank an anonymous referee for a constructive and useful report which helped to improve the paper.  RDA acknowledges support from the Science \& Technology Facilities Council (STFC) through an Advanced Fellowship (ST/G00711X/1).  Theoretical Astrophysics in Leicester is supported by an STFC Rolling Grant.  This research used the ALICE High Performance Computing Facility at the University of Leicester. Some resources on ALICE form part of the DiRAC Facility jointly funded by STFC and the Large Facilities Capital Fund of BIS.


\label{lastpage}

\end{document}